\documentclass[12pt,preprint]{aastex}
\usepackage{epsfig,natbib}
\usepackage{emulateapj5}
\slugcomment{Center Preprint Number: CGPG-03/9-3}
\shortauthors{Bogdanovi\'c et al.}
\shorttitle{Tidal Disruption Signature}

\newcommand{\etal}{{et al.}}

\newcommand{\Msun}{\>{\rm M_{\odot}}}

\begin{document}

\title{Tidal Disruption of a Star By a Black Hole : Observational Signature}

\author{Tamara Bogdanovi\'c\altaffilmark{1}, Michael Eracleous, Suvrath Mahadevan,
Steinn Sigurdsson\altaffilmark{1}, and Pablo Laguna\altaffilmark{1}}

\affil{Department of Astronomy \& Astrophysics, The Pennsylvania State
University, University Park, PA 16802}

\altaffiltext{1}{also member of Center for Gravitational Wave Physics}

\email{tamarab, mce, suvrath, steinn, pablo@astro.psu.edu}

\begin{abstract}
We have modeled the time-variable profiles of the H$\alpha$ emission
line from the non-axisymmetric disk and debris tail created in the
tidal disruption of a solar-type star by a $10^{6}\Msun$ black
hole. Two tidal disruption event simulations were carried out using a
three dimensional relativistic smooth-particle hydrodynamic code, to
describe the early evolution of the debris during the first fifty to
ninety days. We have calculated the physical conditions and radiative
processes in the debris using the photoionization code CLOUDY. We
model the emission line profiles in the period immediately after the
accretion rate onto the black hole became significant. We find that
the line profiles at these very early stages of the evolution of the
post-disruption debris do not resemble the double peaked profiles
expected from a rotating disk since the debris has not yet settled
into such a stable structure. As a result of the uneven distribution
of the debris and the existence of a ``tidal tail'' (the stream of
returning debris), the line profiles depend sensitively on the
orientation of the tail relative to the line of sight. Moreover, the
predicted line profiles vary on fairly short time scales (of order
hours to days). Given the accretion rate onto the black hole we also
model the H$\alpha$ light curve from the debris and the evolution of
the H$\alpha$ line profiles in time.
\end{abstract}


\section{Introduction}

\subsection{Tidal Disruption of a Star by a Black Hole and Related Issues}

A star in an orbit around a massive black hole can get tidally
disrupted during its close passage by the black hole. After several
orbital periods the debris from the disrupted star settles into an
accretion disk and gradually falls into the black hole
\citep*{Rees,CLG,SC,LU}. As material gets swallowed by the black hole
intense UV or soft-X ray radiation is expected to emerge from the
innermost rings of the accretion disk
\citep{FR,LS,Frank,Phinney,SW,MT,SU}. For black hole masses $M_{bh} <
10^{7}\Msun$, tidal disruption theory predicts flares with luminosities
of the order of the Eddington luminosity with durations of the order
of months, and with spectra that peak in the UV/X-ray domain band
\citep{Rees,EK,Ulmer,KPL,Gezari02}. High-energy flares from the
central source illuminate the debris, the photons get absorbed, and
some are re-emitted in the optical part of the spectrum (i.e. the
light is ``reprocessed''). One of the spectral lines in which this
phenomenon can be observed is the Balmer series H$\alpha$ line
($\lambda_{rest}=6563$~\AA).

The disruption of a star begins when the star approaches the tidal
radius, $r_{t}\simeq r_{\star}(M_{bh}/M_{\star})^{1/3}$, the point where
the surface gravity of the star equals the tidal acceleration from the
black hole across the diameter of the star ($r_{\star}$ and
$M_{\star}$ are the radius and mass of the star and $M_{bh}$ is the
mass of the black hole). A $10^{6}\Msun$ black hole is often used as a
prototypical example in tidal disruption calculations. This choice is
motivated by the criterion for a solar-type star to be disrupted
before it crosses the black hole horizon (i.e. the Schwarzchild
radius, $r_{s}$) in order for emission to be observable. For
supermassive black holes with $M_{bh} > 10^{8}\Msun$, $r_{s}>r_{t}$ and the
star falls into the black hole before it gets disrupted.

The tidal disruption process has been the subject of many simulations
\citep*{CL82,CL83,BG,EK,KNP,Laguna,Frolov,MLB,Deiner,ALP,IN,ICN}. It
has been shown that tidal processes in the vicinity of a massive black
hole could lead to tidal capture, tidal heating and tidal spin-up of a
star \citep*{NPP,AK,AH,AM}, and in some cases ultimately to the
explosion of the star. Such explosions, as well as accretion of
post-disruption debris, should manifest themselves as luminous flares
from the centers of galaxies \citep{CL82,Rees}.  Stars close to a
black hole may experience mixing or may eject some of their mass
\citep{Alexander}. As a consequence, stellar populations in nuclear
clusters are expected to be somewhat unusual in comparison with
populations whose evolution was not affected by a massive black hole
\citep{AL,DiStefano}. This has important implications for observations
of stellar cluster in the center of our Galaxy
\citep*{Ghez,Schoedl,EOG,Figer,Gezari} where high concentration of
otherwise rare blue He supergiants has been observed
\citep{Krabbe,Najarro}.

The tidal disruption and accretion of stars can fuel black holes in
the centers of galaxies \citep{Hills,DS,DDCa,DDCb,MCD,FB,Yu} and its
contribution to nuclear activity in galaxies and the growth of the
black hole mass depends on the rate of disruption events in a
galaxy. The predicted tidal disruption rate in a typical inactive
galaxy is $10^{-4}$--$10^{-5}~{\rm yr}^{-1}$ \citep{MT,Alexander}.
This value is consistent with the rate of UV/X-ray outbursts observed
with {\it ROSAT} from inactive nuclei selected as tidal disruption
candidates \citep{Donley}. The rate of tidal disruptions in active and
more luminous nuclei is estimated to be lower, with the lowest value
of $10^{-9}\,{\rm yr}^{-1}$, for galaxies with the most massive black
holes. This may occur partly because massive central black holes
($M_{bh} >10^{8}\Msun$) swallow stars promptly, without disruption,
and partly because stars are less centrally concentrated in these
galaxies \citep{MT}. The observed UV/X-ray flaring rate in these galaxies is
about $9\times10^{-4}\,{\rm galaxy}^{-1}\,{\rm yr}^{-1}$
\citep{Donley} suggesting that in such nuclei, outbursts
may be due to another mechanism, such as accretion-disk instabilities
\citep*{SCK,BKS}.

The tidal encounter of a compact star with a black hole can also
result in emission of gravitational waves, which may be observable
with upcoming instruments. More specifically, compact stars (helium
stars, white dwarfs, neutron stars, and stellar-mass black holes)
which can withstand large tidal forces without being disrupted, may
get captured in relativistic orbits around a supermassive black
hole. Due to the in-spiral and decay of the orbit those objects are
expected to emit the peak of their gravitational wave power in the
LISA frequency band \citep{HB,SR,Freitag01}. It has been recently
suggested by \citet{Freitag03} that very-low mass main sequence stars
(MSSs; $M\ll 1~{\rm M}_{\odot}$) may contribute to events detected by
LISA, which was not previously expected for capture of these objects
by a supermassive black hole. Although MSSs produce a relatively weak
gravitational signal during the in-spiral, compared to compact
objects, their detection in the Galactic center is more likely because
such stars have a predicted close-encounter rate that is an order of
magnitude higher than that of white dwarfs (WDs), and about two or
more orders of magnitude higher than that of neutron stars (NSs) and
stellar-mass black holes (BHs). These compact MSSs are expected to
produce a strong enough signal to allow for 0.5--2 detections from our
Galactic center, with a signal-to-noise ratio of 10 or higher, for a
LISA mission duration of one year \citep{Freitag03}.  Moreover, MSSs
are expected to be disrupted relatively early during the in-spiral,
giving rise to possibly detectable electromagnetic flares. The sudden
appearance of an electromagnetic counterpart to a transient
gravitational wave source, expected in the case of MSSs and helium
stars, could allow identification of the progenitor. In the case of a
tidal disruption event in another galaxy, the coincidence of an
electromagnetic flare and a gravitational wave signal would provide an
indication that the event occurred at the very nucleus of the galaxy,
and possibly allow the measurement of the redshift. More compact
objects which can spiral in without being disrupted (such as WDs, NSs,
and BHs) are expected to produce stronger gravitational wave
signatures and weaker or no electromagnetic flares, with the exception
of white dwarfs in which tidal interaction may trigger thermonuclear
explosion
\citep*{GBW}.

\subsection{Observational Motivation: Transient Emission
Lines in Inactive Galaxies and LINERs}

In view of the above theoretical considerations, it is necessary to
make predictions of the likely observational signatures of a tidal
disruption event. The prompt UV/soft-X-ray flash that is expected to
accompany the disruption does provide strong evidence for such an
event and has in fact been detected in a number of galaxies with {\it
ROSAT} and {\it HST}
\citep*{BPF,Grupea,Grupeb,Donley,BKD,KG,Greiner,GTL,Renzini,LNM}.
However, the duration of this flash is short enough that it can easily
be missed. Aftereffects with a longer duration, such as line emission
from the debris, have a better chance of being detected. If the
appearance of emission lines just after an X-ray flare could be
detected from the same object it could be used to identify the tidal
disruption in the early phase and would provide strong support for the
overall picture, but such cases are rare
\citep*[see][]{Cappellari,Gezari}.

A set of tantalizing observations in the past decade show that several
LINERs \citep[low-ionization nuclear emission regions;][]{Heckman}
have transient Balmer emission lines, which are often double-peaked;
examples include NGC~1097 \citep*{SBBW}, M81 \citep{Bower}, NGC~4450
\citep{Ho}, NGC~4203 \citep{Shields} and NGC~3065 \citep{EH2001}. Such
line profiles are characteristic of rotating disks and resemble the
persistent double-peaked Balmer lines found in about 10--20\% of
broad-line radio galaxies and about 3\% of all active galaxies
\citep[e.g.,][]{EH1994,EH2003,strateva03}. Their abrupt appearance
in LINERs led to suggestions that this transient event was related to
the tidal disruption of a star by a supermassive, nuclear black hole
\citep{ELHS,SB95} or a change in the structure of the accretion disk
associated with a change in accretion rate \citep{SB97}.

To investigate the possibility of line emission from the
post-disruption debris and to evaluate the suggestion that the
transient double-peaked lines of LINERs are related to tidal
disruption events, we have undertaken a calculation of the strength
and profile of the H$\alpha$ line emitted from the debris. In \S2, We
describe two SPH (Smoothed Particle Hydrodynamics) simulations of
tidal disruption on which we base our further calculation of the line
properties.  Our line profile calculation follows the method used for
line profiles emitted by relativistic, Keplerian disks and is
described in \S3. In \S4 we present the resulting line profiles and in
\S5 we discuss the physical conditions in the debris and the
approximations used. In \S6 we summarize our conclusions and consider
future prospects.

\section{Qualitative Description of Two SPH Simulations\label{S_sphsim}}

Tidal disruption simulations were carried out with a three dimensional
relativistic SPH code in order to study the dynamical evolution of the
post-disruption debris. The SPH code used provides a description of
relativistic fluid flows in a static curved spacetime geometry
\citep{Laguna1,Laguna}. We use it to simulate the tidal disruption of a
star in the potential of a Schwarzchild black hole.

The main sequence star is modeled as a polytrope with index
$\Gamma=5/3$.  The density profile of the pre-disruption star is
determined by the Lane-Emden equations. The star is initially placed
on a parabolic orbit at a distance of $700\,r_{g}$ from the black hole
where $r_{g}=r_{s}/2=G M/c^{2}=M$ is the gravitational radius and $M$
is the mass of the black hole.  Hereafter we use units in which
$G\equiv c\equiv 1$, where $G$ is the gravitational constant and $c$
is the speed of the light and we adopt $r_{g}=M$ as a natural unit of
length. The gravitational radius of a $10^{6}\Msun$ black hole is
$r_{g}=1.48\times10^{11}~{\rm cm}=4.92$ light seconds and the
dynamical time at a given radius is $\tau_{dyn} \sim
(r^{3}/GM)^{1/2}=4.92\,(r/r_{g})^{3/2}$~s.

The strength of the tidal encounter, is given by the ratio of tidal
radius to pericentric distance, $\eta$=$r_{t}/r_{p}$ . For the case of
$1\Msun$ star and a $10^{6}\Msun$ black hole $r_{t}\simeq 47\; r_{g}$.
The two SPH simulations describe the tidal disruption for the case of
mildly relativistic encounter, $\eta$=1.2. This value has been
selected as a likely scenario for tidal disruption of a main sequence
star. We do not investigate the strongly relativistic cases where
tidal compression could lead to the explosion of a star, because such
an explosion could introduce an uncertainty in the distribution of the
debris mass over binding energy and consequently in the spatial
distribution and kinematics of the debris.

The self gravity of the star is accounted for initially. Once the star
gets disrupted the self-gravity becomes unimportant and the debris
particles follow nearly Keplerian orbits. It can be shown for the case
of the Keplerian potential that the rate of return of bound debris to
the pericenter follows $dM/dt \propto t^{-5/3}$ \citep{Rees,Phinney}.
This behavior of the debris return rate has been observed in our SPH
simulations. Once bound debris starts to rain down on the black hole
it is expected to cause the initial rapid rise in the emitted UV/X-ray
light curve and steady decay with the power law index of $-^5\!/\!_3$
later on.

The two different simulations have 5,000 and 20,000 particles
(hereafter 5k and 20k respectively) contributing equally to the mass
of a $1\Msun$ star. The 5k SPH calculation follows the debris for 94
days in total. After 34 days significant accretion onto the black hole
begins. Our investigation follows the evolution of the line profiles
in the last 60 days. The 20k SPH simulation spans 53 days, during
which the evolution of the line profiles was followed for the last 6
days (Table 1). Using both the 5k and 20k SPH simulations in the line
profile modeling we take advantage of the longer time span in the
former and better resolution achieved with the larger number of
particles in the latter.

\begin{figure}
\epsscale{0.5}
\plotone{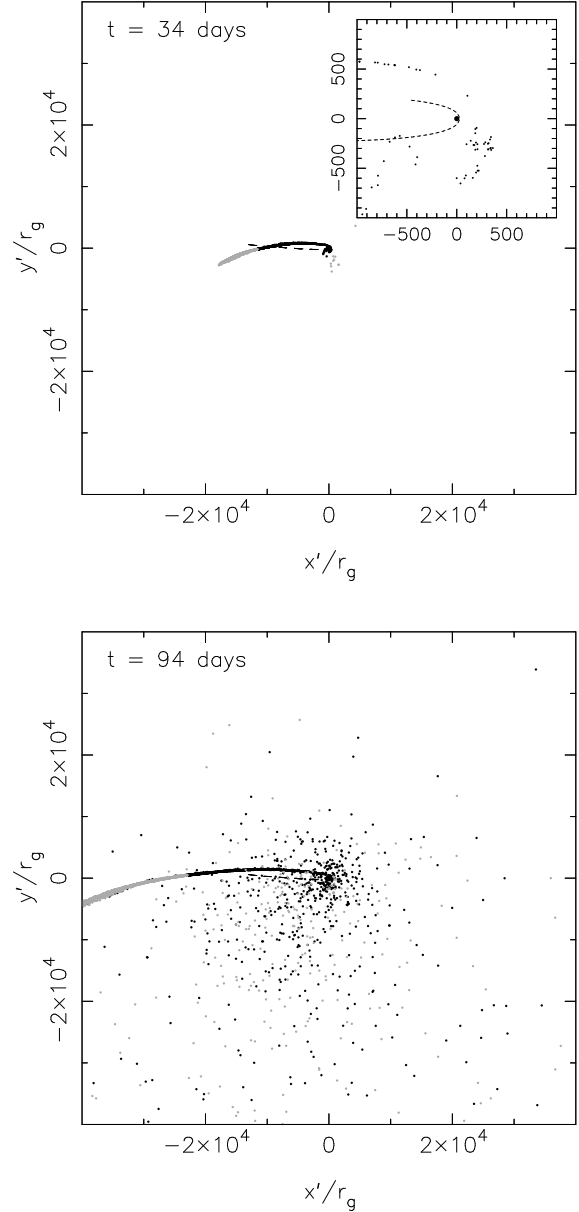} 
\figcaption[f1.eps]{Maps showing the positions of the SPH particles
from the 5k simulation after the second pericentric passage, projected
in the x$'$y$'$-plane at two different times. Particles
gravitationally bound to the black hole are colored black, while the
unbound particles are colored grey. The dashed line represents the
initial trajectory of the star before disruption and the trajectory of
the center of mass of the debris after disruption. The maximum
particle velocities are of order $10^{-2}c$. {\it Upper Panel:}
Particle map at the start of the accretion phase, 34 days after the
disruption occurred. Inset: Particles in the inner region of the
debris, orbiting close to the black hole.{\it Lower Panel:} Particle
map at the end of the simulation, 94 days after the disruption. The
tidal tail can be clearly separated into particles that are unbound
and about to escape the black hole and particles that are returning
towards the black hole. The inner region of the debris consists of
returning particles from the inner tail that have been scattered and
form a rotating structure around the black hole.
\label{fig_map}}
\end{figure}

Figure~\ref{fig_map} shows particle distribution maps after the second
pericentric passage, at the beginning of the accretion phase and at
the end of the 5k simulation. At the early stages of the tidal event
most of the particles were located in the pronounced tidal tail. Sixty
days later, about $20\%$ of the particles are scattered from the tidal
tail and form a quasi-spherical distribution, with most of its mass
concentrated in the equatorial plane \citep{CLG,LU,UPG,Ulmer,MQ}. This
is a consequence of the intersection of the leading part of the tidal
stream with itself \citep{Kochanek,LKR,KPL,ALP}. We refer to the
spheroidal part of the debris as the halo and to its planar component
as the disk. The remaining $79\%$ of particles are still confined to
the tail and $1\%$ are accreted onto the black hole. There is a
concern that some fraction of particles
($\sim\frac{\tau_{run}}{\tau_{dyn}}\sqrt{N}$, where $\tau_{run}$ and N
are the total duration and total number of particles in the run) of
the halo are an artifact of the SPH simulation. This can be due to the
tendency of the SPH numerical method to preserve the constant number
of neighbor particles for each particle during the calculation. In the
regions with a small density of particles this leads to a
``smoothing'' over a large spatial range and it may introduce the
scatter of particles from the debris plane to the halo. These
particles cannot be distinguished from the population of particles
scattered out of the debris plane by the intersection of the tidal
tail with itself. We further discuss the implications of the
spheroidal halo for the emission line profiles and total H$\alpha$
luminosity in
\S\ref{S_halo}.

The velocity distribution in the tail is ``bimodal'' where the central
part of the tail exhibits very low radial velocities: particles on the
near (right-hand) side of the tail ``flow'' towards the black hole,
while particles on far (left-hand) side are moving in the opposite
direction. This is a consequence of the energy distribution throughout
the debris in the disruption process: after the initial disruption
event $50\%$ of the debris stays bound to the system and the other
$50\%$ is unbound. This effect has been predicted by theory in the
case of stellar disruption after a single fly-by of the star
\citep{Rees} and has been observed in tidal disruption
simulations. The symmetry in the distribution of the debris over
binding energy is a consequence of the spin-up of the star at the
expense of orbital kinetic energy. The spin-up initially causes the
development of a quadrupolar deformation. As the tidal interaction
gets stronger the star starts to shed its mass since the material in
the stellar bulge has reached the escape velocity at the star's
surface. One portion of the stellar debris ends up deeper in the
potential well of the black hole, which causes further spread in
binding energy of the debris. This effect determines which portion of
the debris stays bound to the black hole \cite{Rees}. Following the
second passage of the debris through pericenter approximately 66$\%$
of the mass is unbound, 33$\%$ remains bound and only about 1$\%$ is
accreted by the black hole. The maximal approaching and receding
velocities in the debris, with respect to the stationary observer
positioned at infinity, are of order $10^{-2}$c. The dynamical
evolution in the 20k run is the same, it is just followed over a
shorter evolutionary time scale. Since the tidal tail includes a large
fraction of bound and unbound particles in both simulations, its
morphology and velocity field greatly influence the observed line
profiles.

\section{Line Profile Calculation and Time Delay of Reprocessed Light}

\subsection{Calculation of Line Profiles\label{S_prof}}

\begin{figure}
\epsscale{0.5}
\plotone{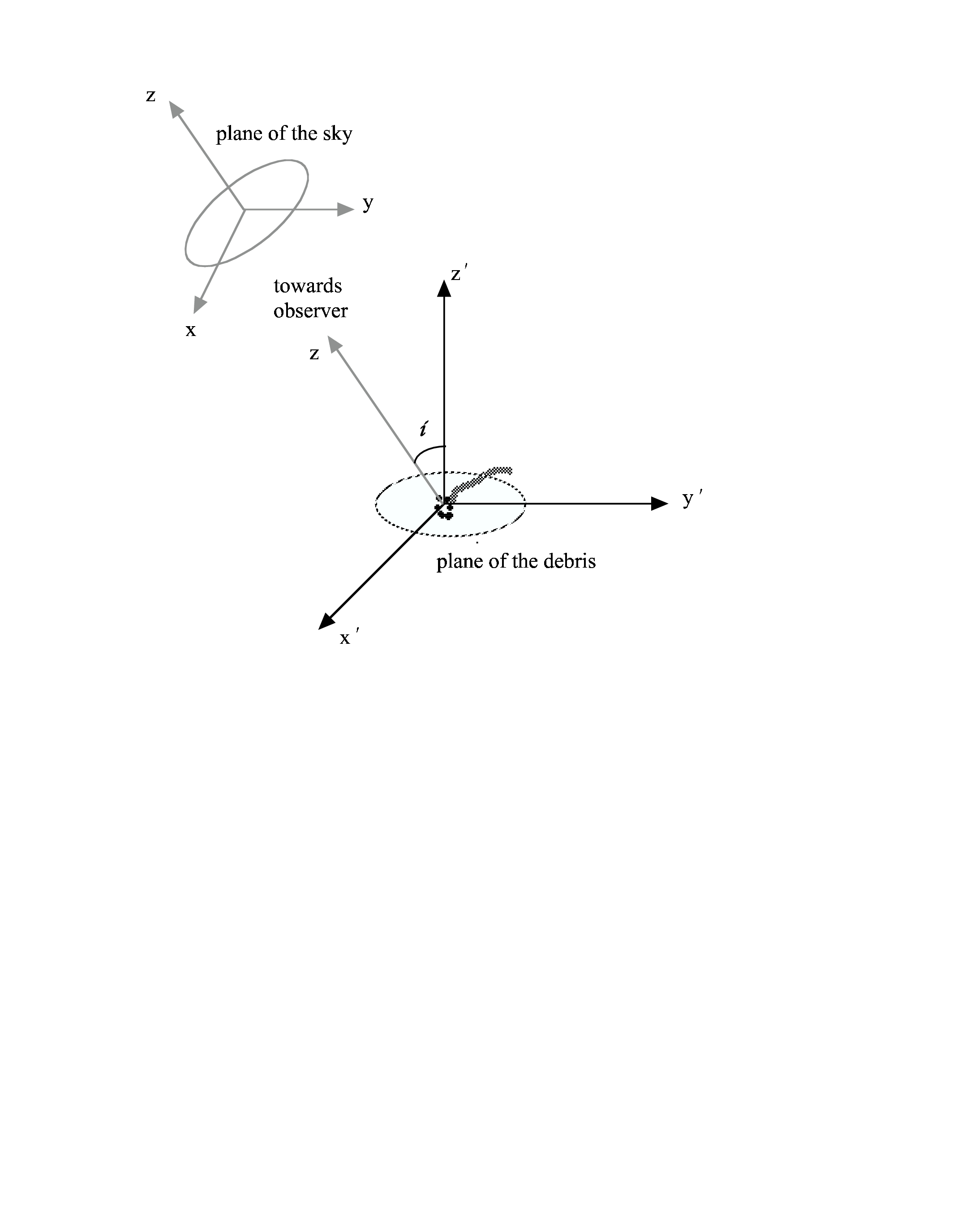}
\figcaption[f2.eps]{The general scheme of the debris geometry used
in the profile calculations. The debris field lies within the
x$'$y$'$-plane, while the observer is located at the infinity in the
direction of the z-axis, with a line of sight that makes an angle $i$
to the z$'$-axis. \label{fig_sketch}}
\end{figure}

We follow the line profile calculations carried out by \citet{CH} and
\citet{ELHS} to obtain the observed profile from a Keplerian,
relativistic, thin disk in the weak field approximation. The
description of the debris as a flat, thin structure is justified by
the fact that the height of the debris is three orders of magnitude
less then its dimensions in the orbital plane. The main objective of
the calculation is to obtain the final expression for the flux density
in the observer's frame as a function of parameters defined in the
reference frame of the debris. Figure~\ref{fig_sketch} shows the
coordinate system and the geometry of the debris. The observer is
located on the positive z-axis, at a distance $d\to +\infty$ and above
the orbital plane at $i=30^{\circ}$ to the z$'$-axis. Since the
calculation is presented in cited papers we just introduce its main
steps and comment on its application to the case of tidal disruption
debris.

The total emission-line flux received from the debris by an observer
at infinity is given by an integral over the plane of the image
produced at infinity, namely
\begin{equation}
F=\int d\nu \int \int d\Omega\, I_{\nu}
\label{eq_F}
\end{equation}
where $\nu$, $I_{\nu}$, and $d\Omega$ are the frequency, specific
intensity and solid angle element measured in the frame of the image
(i.e., of the observer). Using the impact parameter of rays at
infinity, $b$, as a coordinate in the image plane and exploiting the
Lorentz invariance of the quantity $I_{\nu}/ \nu^{3}$ and the fact
that the debris is confined to a plane, equation (\ref{eq_F}) can be
transformed into an expression for the flux density (i.e., the line
profile) in terms of coordinates and physical quantities in the source
frame
\citep*[see detailed derivation in][Chen \& Halpern 1989, and
Eracleous et al. 1995]{CHF}
 $$
 f_{\nu}=  \int\int d\Omega\, I_{\nu} = \hbox to 2 truein{\hfill}
 $$
\begin{equation}
\frac{M^{2}\nu_{0}\cos i}{d^{2}}\
\int_{\xi_{in}}^{\xi_{out}} \!\!\!\!\xi\, d\xi \,\int_0^{2\pi} \!\!d\phi'\;
I_{\nu_{e}}\; D^{3}(\xi,\phi')\;\psi(\xi,\phi') 
\label{eq_Fnu}
\end{equation}
where the new polar coordinates in the debris plane are the
dimensionless radius $\xi\equiv r/r_{\rm g}$ and the azimuthal angle
$\phi'$. In practice, the integration is performed by summing over
particles, assigning to each particle an emissivity according to its
position, as derived from calculation with the code CLOUDY
\citep[][see \S\ref{S_lcurve}]{Ferland}. The limits of integration 
describe the portion of the debris that emits the H$\alpha$ line,
between radii $\xi_{in}$ and $\xi_{out}$.  The function
$D(\xi,\phi')$, the ``Doppler factor,'' is determined by the phase
space distribution of the emitting particles and the metric, and
describes the effects of gravity and the motion of the emitting
particles on the energies of the emitted photons.  The function
$\psi(\xi,\phi')$ is determined by the geometrical distribution of the
emitting particles and the metric, and describes the effects of curved
trajectories of light rays. In the special case where the debris is
confined to a plane, these functions are given by
 $$
 D ={(1-2/\xi)^{1/2}\over\gamma} \left\{1-
 \frac{\beta_{r'}\, \left[1- (b/r)^{2} (1-2/\xi)\right]^{1/2}}{1-2/\xi}
 \right.
 $$
\begin{equation}
\left.\phantom\ + \frac{\beta_{\phi'}\, (b/r)\,\sin i \,\sin \phi'} 
{(1-\sin^{2}i \cos^{2}\phi')^{1/2}} 
\right\}^{-1} ,
\end{equation}
with $\gamma$ the Lorentz factor, $\beta_{r'}$ and $\beta_{\phi'}$ the
radial and azimuthal velocities of debris particles {\it in the source
frame}, and
\begin{equation}
\psi (\xi,\phi')=1+\frac{1}{\xi}\ \left(\frac{1-\sin i \,\cos 
\phi'}{1+\sin i \,\cos \phi'}\ \right) .
\end{equation}
The above analytic expression for $\psi (\xi,\phi')$ has been derived
in the weak field approximation and is accurate to order
$\xi^{-1}$. This approximation is appropriate in our case because the
portions of the debris that experience a strong gravitational field
are also highly ionized and make a negligible contribution to the
H$\alpha$ flux (see discussion in \S\ref{S_phys}).  The ratio $b/r$
describes how rays emitted from the debris are mapped to points in the
image at infinity and is given by $b/r = (1-\sin^{2}i
\cos^{2}\phi')^{1/2}\, \psi (\xi,\phi')$. Finally, the Lorentz factor is
given by
$
 \gamma = [1- {\beta_{r'}^2(1-2/\xi)^{-2}} -
{\beta_{\phi'}^2(1-2/\xi)^{-1}}]^{-1/2}\; .
$
 
The emission properties of the debris are described by the local
specific intensity, $I_{\nu_{e}}$. We take the local line profile to
be a Gaussian corresponding to a velocity dispersion $\sigma$ (in
units of the speed of light).  The width of the local line profile
represents not only internal motions of the line-emitting gas (not
captured by the SPH simulation) but also the velocity range between
the discretized debris points we use in our numerical integration. We
further assume that the emissivity of the line is a power-law with
radius (see discussion in \S\ref{S_lcurve} and \S\ref{S_phys}).
Therefore, we write the specific intensity as
\begin{equation}
I_{\nu_{e}}=
\frac{\epsilon_{0}\,\xi^{-q}}{2\,(2\pi)^{3/2}\,\sigma} 
\;\exp\left[-\frac{(1+X-D)^{2}}{2\,D^{2}\sigma^{2}}\right]
\label{eq_emiss}
\end{equation}
where $X$ is defined by $1+X\equiv\nu/\nu_0$, where $\nu$ and $\nu_0$
are observed and rest frame frequency, and $\epsilon_0$ is a constant.

In the final model, the line profiles are described by the following
parameters: the inner and outer radius of the line-emitting portion of
the debris, $\xi_{in}$ and $\xi_{out}$, the particle emissivity
power-law index $q^{tail}$ and $q^{disk}$, the inclination of the
debris plane $i$, the local velocity dispersion $\sigma$ in units of
$c$, and the radius of the central continuum source $\xi_{0}$. The
last parameter is relevant to the calculation of the light travel time
across the debris, which we describe in
\S\ref{S_delay}, below.

\subsection{Time Delay of the Reprocessed Light Emitted by the Debris\label{S_delay}}

In the emission model we adopt, a central source of finite dimensions
illuminates the debris. The luminosity of this source is proportional
to the accretion rate onto the black hole.  Due to the finite velocity
of light, at any given time (in the observer's frame), different
portions of the debris are seen to respond to a different level of
illumination.  It is also noteworthy that the length scales and time
scales in this problem span a very large dynamic range.  As a
consequence, the light-crossing time of the outer portions of the
debris is longer than the dynamical time of the inner portions, which
makes it necessary for us to follow the redistribution of the debris
in phase space and the variations of the X-ray source carefully.

It is simple to show that travel time delay for light rays can be
written as
\begin{equation}
\Delta t_
{travel}=(\xi-\xi_{0})(1-\sin i \,\cos \phi')
\end{equation}
where $\xi$ and $\phi'$ are the coordinates of a particle in the
orbital plane, $\xi_{0}$ is the radius of the central source and $i$
is the inclination.  Since light rays travel in the gravitational
potential of the black hole they suffer an additional, relativistic
time delay, which can be calculated from the equation of geodesics for
photons
\citep[][p.202]{Weinberg}. The assumptions are that photons travel in
an isotropic gravitational field and that their trajectories can be
considered coplanar with the observer and the black hole. We consider
only the gravitational delay caused by the black hole's potential and
assume that the debris does not have any significant gravitational
influence on a light ray. This is a reasonable assumption since
$\rho_{debris} / \rho_{bh}\sim 10^{-12}$. In our notation, the
gravitational time delay for non-rotating black hole can be expressed
as
 $$
 \Delta t_{gr} \cong 
 \left(\frac{\xi-\xi_{0}}{\xi+\xi_{0}}\right)^{1/2}\!\!
 +2\,\ln\left[\frac{\xi+\left(\xi^2-\xi_0^2\right)^{1/2}}{\xi\cos i}\right]
 \hbox to 0.5truein {\hfill}
 $$
\begin{equation}
%\Delta t_{gr} \cong 
%\left(\frac{\xi-\xi_{0}}{\xi+\xi_{0}}\right)^{1/2}\!\!
%+2\,\ln\left[\frac{\xi+\left(\xi^2-\xi_0^2\right)^{1/2}}{\xi\cos i}\right]
-\cos \phi'\left[\left(\frac{1-\cos i}{1+\cos i}\right)^{1/2}\!\!
+2\,\ln\left(\frac{1+\sin i}{\cos i}\right)\right] .
\end{equation}
The overall delay for a particle caused by light travel and
general relativistic effects is then
\begin{equation}
\Delta t=\Delta t_{travel}+\Delta t_{gr} .
\end{equation}
We only need to account for relative time delays, which we calculate
relative to a ray coming from the origin of the coordinate system of
the debris. We find that the general relativistic time delay in our
calculations never exceeds $10\%$ of the travel time delay, and it is
typically of order a few percent.

\section{Results and Implications}

\subsection{Light Curves and Emissivity of the Debris \label{S_lcurve}}

The output of the 5k run comprises 351 frames describing the evolution
of the debris morphology over 60 days of accretion, with a time step
of approximately 4 hours (see Table 1). This is a fine enough temporal
resolution to trace the redistribution of particles in the debris. For
comparison, a particle at $r=200\,r_{g}$ orbits the black hole with a
period of approximately 24.5 hours.  During the accretion phase of the
simulation the number of particles decreases due to infall in the
black hole. Tracking this number allows us to follow the accretion
rate and construct the X-ray light curve. The total amount of mass
accreted in this run is about $10^{-2}\,\Msun$. The X-rays resulting
from accretion illuminate the debris out to large distances from the
black hole and power the emission of H$\alpha$ photons.

The output of the 20k run consists of 204 temporal frames spanning 6
days of accretion with a time step of about 45 minutes (the coverage
is not even; it includes gaps since the behavior of the debris can be
captured even with sparse sampling).  Because of the short time span
of accretion, the number of accreted particles is small (less than
0.1$\%$ of the total mass) and consequently, the illuminating,
X-ray light curve is not smooth. To address this issue we compute the
H$\alpha$ light curve for two more, fiducial, illumination light
curves: one consistent with the debris return rate predicted by theory
\citep[$\propto t^{-5/3}$;][]{Rees,Phinney}
\footnote{We have verified that the energy distribution of the debris
particles in our simulation would lead to a $\propto t^{-5/3}$ return
rate of particles to the pericenter, based on the Keplerian orbits},
and the other, constant in time. This allows us to also examine the
effect of different illumination patterns on the line profiles, which
we describe in \S\ref{S_variability}.

In order to calculate the resulting H$\alpha$ luminosity of the debris
it is necessary to determine the efficiency with which the debris
reprocesses the incident radiation. This efficiency can be
characterized by the surface emissivity of the debris as a function of
radius, $\epsilon=\epsilon_{0}\,\xi^{-q}$, as shown in equation
(\ref{eq_emiss}). We used the photoionization code CLOUDY
\citep{Ferland} for numerical calculations of the response of the 
debris to illumination. From the calculated surface emissivity as a
function of radius (${\rm erg\,s^{-1} cm^{-2}}$) we were able to
determine the particle emissivity as a function of radius
$Q=\epsilon/\Sigma=Q_{0}\,\xi^{-\beta}$ (${\rm erg\,s^{-1}
particle^{-1}}$), where $\beta=q-p$ is the particle emissivity
power-law index, given the surface density distribution of the
debris, $\Sigma\propto\xi^{-p}$.

The response generally depends on a spectral
energy distribution (SED) of the incident radiation, matter density
distribution, and the system geometry. We have described the debris
system in terms of these physical parameters and performed CLOUDY
calculations.

We have assumed a SED for the illuminating radiation of the form
$L_{\nu}\propto \nu^{-1}$, extending to 100~keV. The normalization for
the SED is set by the accretion luminosity of the debris. The mass
accreted during the accretion phase of the 5k simulation gives rise to
a time-average accretion luminosity of $L_{acc}\approx 1.5\times
10^{43}~M_6~{\rm erg~s}^{-1}$, while the maximum luminosity achieved
at the beginning of the accretion is $L_{acc}^{max}\approx 8.0\times
10^{44}~M_6~{\rm erg~s}^{-1}$ (where $M_6=M_{bh}/10^6\,{\rm
M_{\odot}}$). The luminosity during flares therefore may exceed the
Eddington luminosity, $L_{Edd}=1.51 \times 10^{44}~M_6~{\rm
erg~s}^{-1}$. For the purpose of CLOUDY calculations, in order to
characterize the emissivity of the debris, we have used the
time-average value for the accretion luminosity. Our estimated
accretion luminosity is comparable to what was observed by
\citet{Gezari} for three tidal disruption candidates, namely
$L_{X}=10^{42}$--$10^{44}~{\rm erg~s}^{-1}$.

Based on the particle distribution in the simulation we find that the
density in the tidal tail decreases with time because the tail gets
stretched as the debris evolves (cf., Figure~\ref{fig_map}). We find
that for the typical SPH time frame the density distribution in the
tail can be approximated as $n_{H}^{tail}\propto \xi$ and is in the
range $n_{H}^{tail}=10^{14}$--$10^{15}~{\rm cm}^{-3}$. With path
lengths of $9\times10^{13}~{\rm cm}$ and $\sim5\times10^{15}~{\rm
cm}$, the corresponding tail column densities are in the range
$N_{H}^{tail}=10^{27}$--$10^{31}~{\rm cm}^{-2}$. The low and high
values of the column density correspond to lines of sight along the
short and long axes of the tail, respectively. (See
Table~\ref{T_debris} for values of physical parameters in the debris.)
The density of particles scattered from the tail in the spheroidal
halo around the black hole is fairly low in comparison. The halo
density reaches a maximum in the plane of the debris ($\geq3\sigma$
over-density), where particles orbiting around the black hole form a
disk of radius $\sim2500\,r_{g}$. The number density in the disk
decreases with radius as $n_{H}^{disk}=1\times 10^{12}~{\rm
cm}^{-3}(\xi/500)^{-2.1}$. The corresponding column densities in the
directions orthogonally and radially through the disk are
$N_{H}^{disk}\approx 7\times10^{20}-3\times10^{25}~{\rm cm}^{-2}$,
respectively. The spheroidal part of the halo, formed from particles
scattered out from the plane of the debris has the estimated number
density of $n_{H}^{halo}=3\times 10^{11}~{\rm
cm}^{-3}(\xi/500)^{-1.4}$ and corresponding column density of
$N_{H}^{halo}\approx 1\times10^{25}~{\rm cm}^{-2}$.

\begin{figure}
\epsscale{0.5}
\plotone{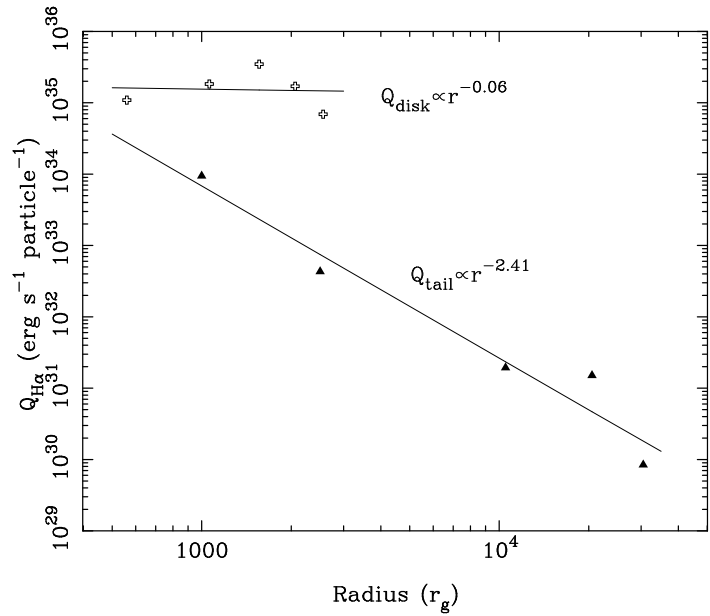}
\figcaption[f3.eps]{The quantum efficiency for particles in the 
disk (upper) and the tail (lower) as a function of distance from the
black hole. The data points are the results of CLOUDY photoionizations
calculations performed for different regions in the tail and the
disk. The solid line is the best fit power law, with power law index,
$q^{tail}$=2.41 and $q^{disk}$=0.06, respectively.
\label{fig_emiss}}
\end{figure}

The CLOUDY calculations show that the H$\alpha$ power emitted by the
mostly neutral tidal tail in response to X-ray illumination decays
with distance from the center as $\epsilon^{tail}\propto r^{-1.6}$, as
a consequence of the density distribution in the tail and the geometry
of the debris (i.e. illumination incidence). The corresponding
particle emissivity for the tail derived from the surface emissivity
decays with the distance as $Q^{tail}\propto r^{-2.4}$
(Fig.~\ref{fig_emiss}).  The disk and the halo components are almost
completely ionized and only 1 in $10^{7}$ hydrogen atoms is neutral,
on average. The emissivity of the disk decreases with distance from
the black hole as $\epsilon^{disk}\propto r^{-2.2}$,
approximately. The equivalent particle emissivity distribution for the
disk is $Q^{disk}\propto r^{-0.06}$. Therefore, in our description of
the surface emissivity of the debris (equation~\ref{eq_emiss}) we use
power-law indices of $q^{tail}=1.6$ for the tail and $q^{disk}=2.2$
for the disk. In our calculation for line profiles and light curves,
due to the discrete nature of the SPH simulation, we use the
prescription for the emissivity per particle, with power-law indices
of $\beta^{tail}=2.4$ for the tail and $\beta^{disk}=0.06$ for the
disk. The results of the CLOUDY calculations and the power-law
prescriptions derived from them are summarized in
Figure~\ref{fig_emiss}. The total $H\alpha$ luminosities contributed
by the tail and the halo for the time-average value of the
illumination $L_{acc}=1.5\times 10^{43}~{\rm erg~s}^{-1}$, as
calculated by CLOUDY, are $L_{\rm H\alpha}^{tail}\simeq
1\times10^{36}~{\rm erg~s}^{-1}$ and $L_{\rm H\alpha}^{disk}\simeq
1\times10^{37}~{\rm erg~s}^{-1}$. The calculated value of the
luminosity for the spheroidal component is $L_{\rm
H\alpha}^{halo}\simeq 6\times10^{38}~{\rm erg~s}^{-1}$ (also listed in
Table~2).

We discuss the emission properties of the debris in
\S\ref{S_phys} and their implications for the observability in
\S\ref{S_obs}. Finally, with the above emissivity prescriptions we
calculate the {\it observed} H$\alpha$ luminosity curve of the debris
at a particular time step by computing the time at which the light was
emitted from the debris and by finding the ionizing flux that was
illuminating that location at the time the emission occurred,
according to the light-travel time from the black hole to that
particular region of the debris.

\begin{figure}
\epsscale{0.4}
\plotone{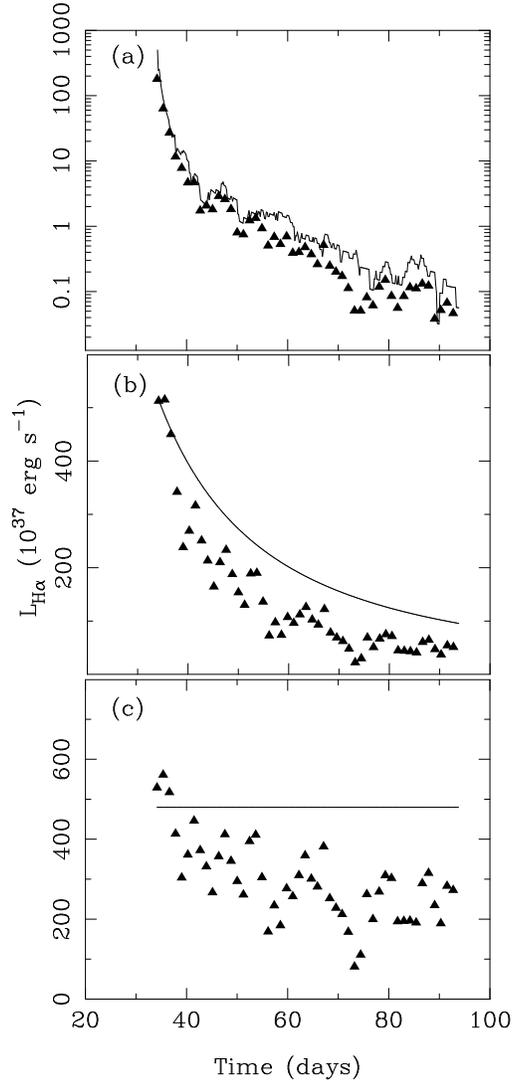}
\figcaption[f4.eps]{The H$\alpha$ light curves (data points)
resulting from the re-processing of three different illumination light
curves. The solid line shows a scaled version of the UV/X-ray
continuum light curve that illuminates the debris, which in a) follows
the SPH accretion rate; b) decays with time as $t^{-5/3}$; c) is
constant. Note that the vertical axis is calibrated logarithmically in
(a), while in (b) and (c) they have the same linear scale. The
H$\alpha$ light curves rise initially as the illumination front
propagates through the debris and then decay faster than the UV/X-ray
light curves. The details of the calculation are described in
\S\ref{S_lcurve}.
\label{fig_lcurve}}
\end{figure}

Figure~\ref{fig_lcurve} shows three different H$\alpha$ light curves
from the debris confined to a plane (assuming $\xi_{in}=500$,
$\xi_{out}=40,000$) during the 60-day accretion phase of the 5k
simulation. Figure~\ref{fig_lcurve}a shows the accretion luminosity on
a logarithmic scale (solid curve), calculated from the accretion rate
of the debris in the SPH simulation. The UV/X-ray luminosity curve is
arbitrarily scaled and overplotted on the top of the H$\alpha$ curve
for comparison. It is noticeable that the H$\alpha$ light curve
departs from the accretion light curve at late times, though the
departure appears small in the logarithmic plot (used here due to the
large dynamic range of the light curves). The same effect is more
noticeable in the Figure~\ref{fig_lcurve}b, where the accretion
luminosity is proportional to $t^{-5/3}$ and the H$\alpha$ light curve
is plotted on a linear scale. The H$\alpha$ light curve roughly
follows the shape of the incident UV/X-ray light curve at early times
but decays faster at late times. The faster decay in the H$\alpha$
light curve reflects the debris evolution in time: as the tail becomes
more elongated, the incident photons travel a longer way to illuminate
the debris. Consequently, the intensity of the illuminating light gets
lower in the later stages of the tidal disruption event. The relative
decay rate of the H$\alpha$ light curve with respect to the UV/X-ray
light curve diminishes about 80 days after the accretion started. We
find that this late relative rise in the H$\alpha$ luminosity is due
to the increase of number of particles in the disk component. As
particles diffuse from the high density tail to the lower density
disk, in later stages of the simulation, their emission efficiency
increases and they contribute a significant amount of H$\alpha$ light
to the light curve. To isolate the effect of the debris evolution in
time from the evolution of the illuminating light curve, we calculate
the H$\alpha$ light curve in the case of constant illumination
(Fig.~\ref{fig_lcurve}c). Here, the relative departure of the
H$\alpha$ light curve from the UV/X-ray light curve can be interpreted
as a consequence of the expansion and redistribution of the
debris. The H$\alpha$ luminosity appears to level off at late times
because the debris disk begins to settle into a quasi-steady
configuration.

In summary, the {\it observed} H$\alpha$ flux depends sensitively on
the UV/X-ray light curve, on the distribution of matter that makes up
the inner portion of the debris, and on how quickly particles
redistribute themselves in phase space. The main features of the
H$\alpha$ light curve are: an initial rise followed by a decline, with
superposed fluctuations. The initial rise is a consequence of the
propagation of the initial illumination front through the debris at
the speed of light. The fluctuations are a result of the fluctuations
in the accretion rate, which are caused, in turn, by the finite number
of particles employed in the simulation. The decay rate of the
H$\alpha$ light curve is determined by the decay rate of the
UV/X-ray light curve, debris expansion and redistribution rate.

\subsection{Line Profiles From the Debris and Their Variability\label{S_variability}}

We have computed sample line profiles emerging from the debris
for the following choices of model parameters:

\begin{enumerate}

\item
{\it Inclination angle of the plane of the debris, $i$. --} We assumed
that the observer is located on the positive z-axis, at a distance
$d\to \infty$ at $i=30^{\circ}$ to the z$'$-axis. Changing the
inclination, changes the values of the projected velocities (i.e., the
overall width of the line profile) but has very little effect on its
shape otherwise.

\item
{\it Inner and outer radius, $\xi_{in}$ and $\xi_{out}$. --} The
adopted inner radius of the debris is the inner boundary of the region
from which H$\alpha$ emission is expected to emerge. The choice of the
inner radius depends on physical conditions in the debris, as we
explain in \S\ref{S_phys}. Here, we explore several cases with
$\xi_{in}$ between 200 and 10,000. The outer radius $\xi_{out}=40,000$
is naturally set by the dimensions of the system.

\item
{\it Particle emissivity power-law index, $\beta$. --} As noted in
\S\ref{S_lcurve}, above, we find that $\beta^{tail}=2.4$ in the debris
tail and $\beta^{disk}=0.06$ in the disk. These indices describe the
emissivity per particle as a power-law and correspond to indices
$q^{tail}$ and $q^{disk}$ in the description of the surface
emissivity. The adopted values for emissivity indices significantly
influence the profile shapes. Higher values of $\beta$ weigh the
emissivity towards smaller radii where the projected velocity is
higher.

\item
{\it Velocity dispersion, $\sigma$, and central source radius
$\xi_{0}$. --} The adopted value of the velocity dispersion for the
profiles presented was $\sigma = 100~{\rm km~s}^{-1}$. A lower limit
on the velocity dispersion is set by the velocity difference measured
for pairs of adjacent particles. This value represents the dispersion
due to the finite number of resolving elements in the simulation and
equals $20~{\rm km~s}^{-1}$ for 90$\%$ of particles close to the
debris plane. The constraint on the upper limit of the velocity
dispersion comes from the velocity dispersion due to small scale
turbulence, $\sigma_{turb}\le(1/2)(L/r)\,V_{Kepler}
\approx 800~{\rm km~s}^{-1}$, where $V_{Kepler}$ is a local circular
velocity, $r$ is a distance from the massive black hole and $L$ is the
smallest dimension of the fluid (i.e. width or thickness of the
tail). Since the real turbulence could be substantially smaller than
the upper limit due to dissipation by internal shocks, we adopt a
value of $100~{\rm km~s}^{-1}$. Larger values of $\sigma$ produce
wider profiles and smooth-out sharp features. The central source
radius is arbitrarily chosen to be $\xi_{0}=200$. It implies a central
source of finite dimensions such as a corona of ionized plasma, or a
vertically extended accretion flow in the innermost parts of an
accretion disk. Its effect on the line profiles is rather small.

\end{enumerate}

\begin{figure}
\epsscale{0.8} 
\plottwo{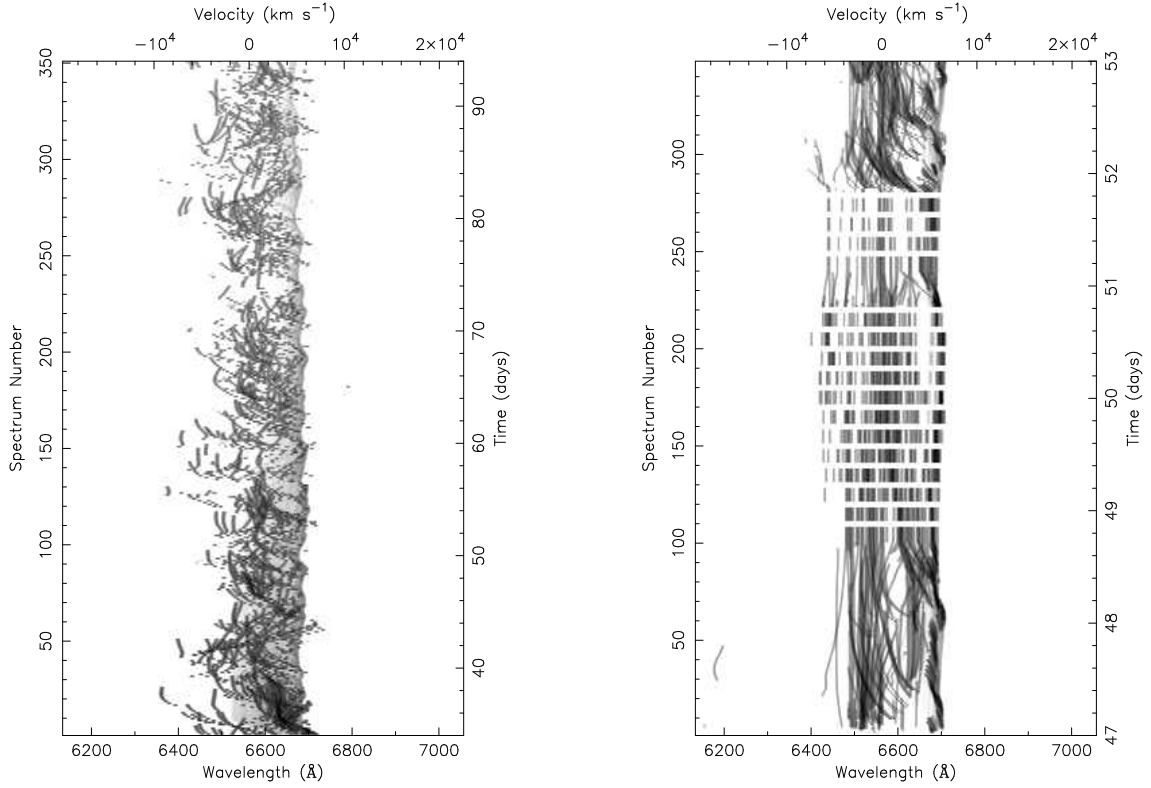}{f5b.eps}
\figcaption[f5.eps]{Trailed spectrogram of the simulated H$\alpha$
emission-line profiles from the 5k simulation spanning 60 days (left)
and from the 20k simulation spanning 6 days (right). This is
effectively a 2-dimensional intensity map versus projected velocity of
the emitting material and time. Darker shades correspond to higher
intensity. The scale on the right represents time since the tidal
disruption event.\label{fig_trail}}
\end{figure}

\begin{figure}
\epsscale{0.7}
\plottwo{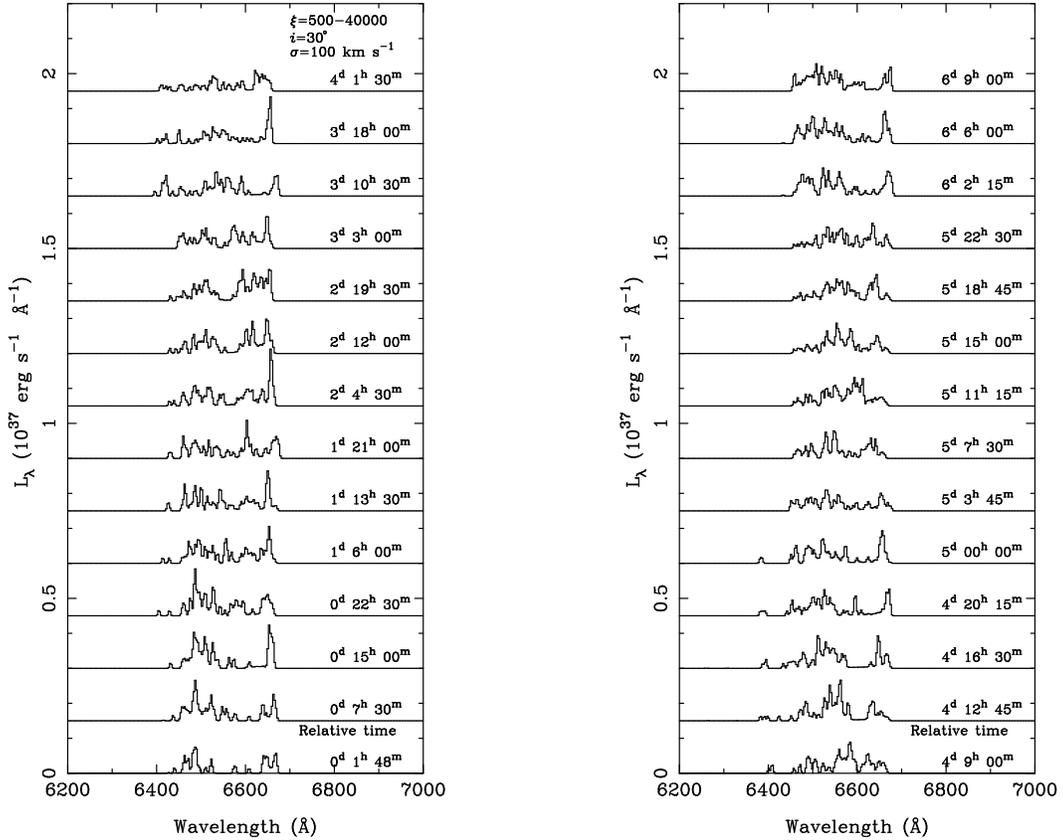}{f6b.eps}
\figcaption[f6.eps]{Sequence of H$\alpha$ profiles emitted from the
region $\xi\in(500,40\,000)$ over period of 6 days (20k run). The
relative time from the beginning of the accretion phase onto the black
hole is marked next to each profile. The accretion phase begins 47
days after the tidal disruption. The inclination of the debris plane
and the velocity shear are as marked on the figure.\label{fig_evol}}
\end{figure}

\begin{figure}
\epsscale{0.4}
\plotone{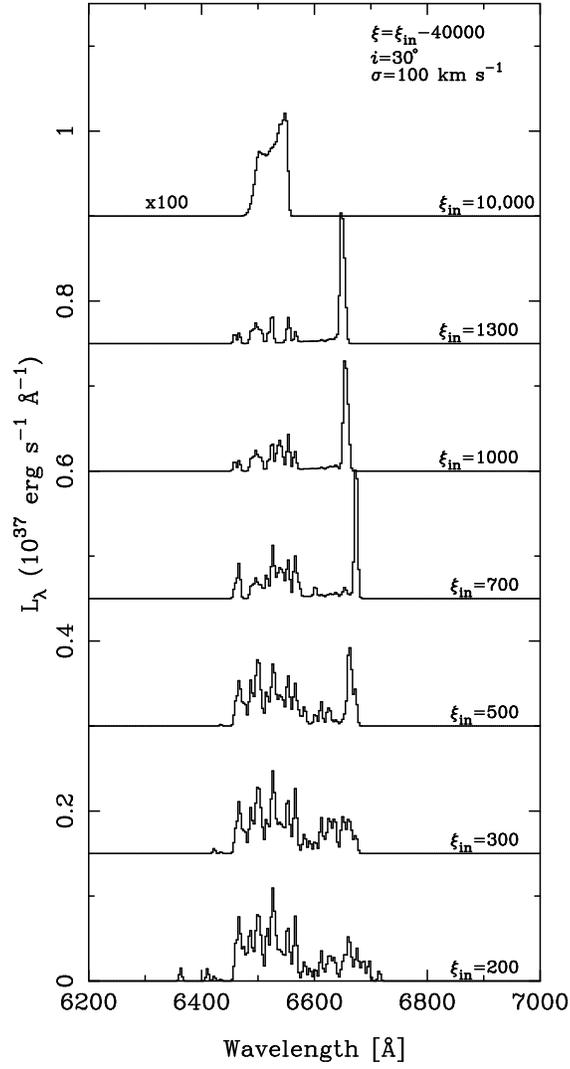}
\figcaption[f7.eps]{H$\alpha$ emission line profiles simulated for seven 
different values of inner radius ($\xi_{in}$), as determined by the
propagation of the ionization front through the debris. The relative
time for profile frames is 6$^{d}$ 6$^{h}$ 0$^{m}$. The intensity of
the profile calculated for $\xi_{in}=10,000$ is multiplied by the
factor of $100$. The inclination and velocity shear are as marked on
the top of the figure.\label{fig_rad}}
\end{figure}

\begin{figure}
\epsscale{0.4}
\plotone{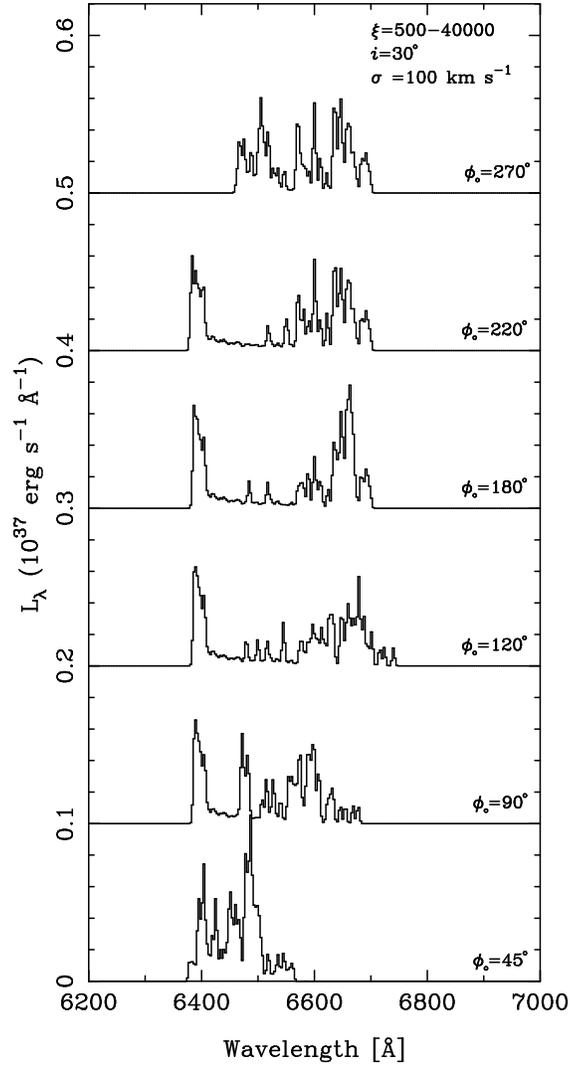}
\figcaption[f8.eps]{H$\alpha$ emission line profiles simulated for six 
different azimuthal orientations of the debris with respect to the
observer, as represented by $\phi_{0}$. See profile in
Figure~\ref{fig_evol} with the time label 5$^{d}$ 22$^{h}$ 30$^{m}$
for orientation $\phi_{0}=0^{\circ}$. The size of the emitting region,
inclination and velocity shear are as marked on the top of the
figure.\label{fig_ori}}
\end{figure}

\begin{figure}
\epsscale{0.4}
\plotone{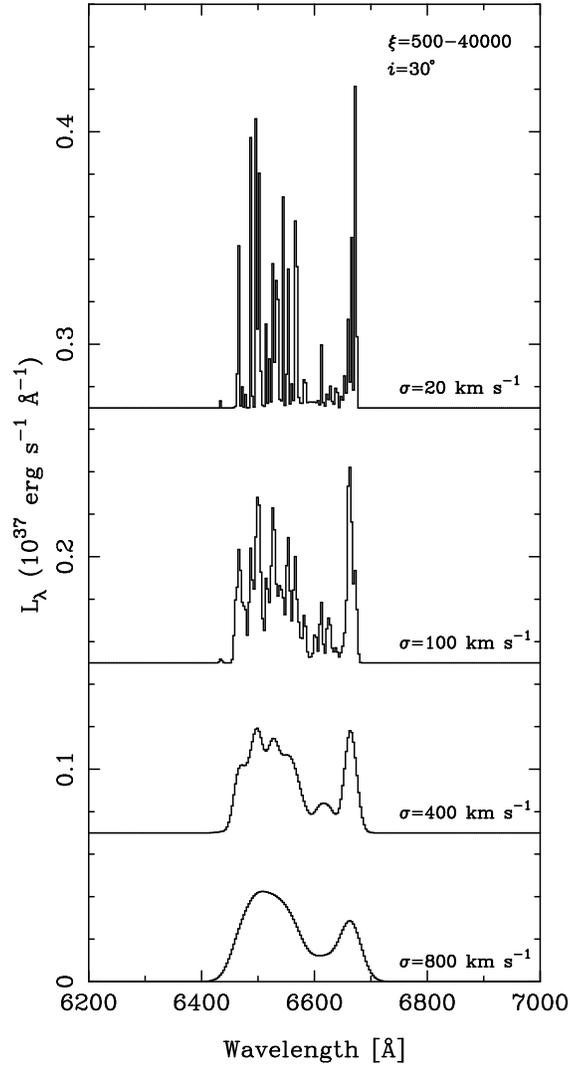}
\figcaption[f9.eps]{H$\alpha$ emission line profiles simulated for 
four different values of velocity dispersion. A velocity dispersion
of 100$\,km\,s^{-1}$ is the equivalent of an instrumental resolution
of 2~\AA. The relative time for the profile frames is 6$^{d}$ 6$^{h}$
0$^{m}$. Size of the emitting region and inclination are as marked on
the top of the figure.\label{fig_sig}}
\end{figure}

In Figures~\ref{fig_trail}--\ref{fig_sig} we show sample line profiles
to illustrate how they evolve in time and how they are affected by the
choice of model parameters and by the orientation of the observer.
Figure~\ref{fig_trail} is a ``trailed spectrogram'' summarizing the
temporal evolution of the line profiles from the two different SPH
runs; it is a 2-dimensional map of the H$\alpha$ emission as a
function of projected velocity and time. Figure~\ref{fig_evol} shows a
different representation of the evolution of the line profile with
time, which effectively comprises selected time slices from the
trailed spectrogram. Figures~\ref{fig_rad} and \ref{fig_ori} show how
the inner radius of the line-emitting region and the azimuthal
orientation of the observer affect the observed line
profiles. Figure~\ref{fig_sig} shows the effect of the different
values of velocity dispersion on the shape of the line profiles. The
main properties and features of our results are as follows:

\begin{description}

\item[{\it Profile variability with time. --}] A property that is 
immediately obvious in the line sequence is the change of the profile
shape with time (Figures~\ref{fig_trail} and \ref{fig_evol}). It is
noticeable that the adopted low value of velocity dispersion allows us
to resolve individual particles in the trailed spectrograms, orbiting
around the black hole. The evolution of the line intensities in time
roughly follows the behavior of the UV/X-ray luminosity but decays
somewhat faster in time (see \S\ref{S_lcurve}). The multi-peaked line
profile is a consequence of the velocity field of the inner debris,
which consists of the inner portion of the tidal tail that is falling
towards the black hole (towards the observer) and debris that is
rotating around the black hole after being scattered.  The line
profiles and their variability could be observationally important
features of the debris just formed from tidal disruption. The variable
line profiles might be observed and recognized on the relatively short
time scale of hours to days.

\item[{\it Effect of the inner radius, $\xi_{in}$. --}] The profiles
become broader as the inner radius of the line-emitting regions
decreases since higher-velocity gas resides at smaller radii (see
Figure~\ref{fig_rad}). The approximate full width at zero intensity of
the profiles ranges from $4,500~{\rm km~s}^{-1}$ for $\xi_{in}=10,000$
to $18,000~{\rm km~s}^{-1}$ for $\xi_{in}=200$. We find that line
profiles change from the profiles dominated by the emission red-ward
from the rest wavelength for $\xi_{in}<1000$ to narrower profiles
dominated by the blue-ward emission from the tail for $\xi_{in}>1000$,
since for large values of $\xi_{in}$, the high-velocity rotating gas
in the vicinity of the black hole is excluded and the dominant
contributions to the line profile come from the tidal tail. The
intensity of the line also decreases with increasing inner radius,
making the outer regions of the debris harder to observe.

\item[{\it Effect of observer orientation, $\phi_{0}$. --}] Because of
the non-axisymmetric geometry and velocity field, the line profiles
emitted by the debris, depend on the orientation of the tidal tail
relative to the observer. In Figure~\ref{fig_ori} we show the effect
of azimuthal orientation $\phi_{0}$ of the debris, with respect to the
observer. The values of $\phi_{0}$ are $45^{\circ}$, $90^{\circ}$,
$120^{\circ}$, $180^{\circ}$, $220^{\circ}$ and $270^{\circ}$, as
measured in a counterclockwise direction from positive x$'$-axis to
the observer's line of sight. These can be compared with the profile
corresponding to the same time in Figure~\ref{fig_evol} for
$\phi_{0}=0^{\circ}$. The position of the peaks in
Figure~\ref{fig_ori} varies relative to the rest wavelength, since the
relative direction of bulk motion of the material depends on the
observer's orientation. For example, it is possible to distinguish the
emission from the tail for the range of azimuthal orientations
$90^{\circ}-220^{\circ}$. The tail emission in these profiles appears
as the most blueshifted peak, since these are the orientations for
which different portions of the tail flow towards the observer.

\item[{\it Effect of velocity dispersion, $\sigma$. --}] In 
Figure~\ref{fig_sig} we show the effect of four different values of
the velocity dispersion in calculation of the emission line
profiles. A value of $20~{\rm km~s}^{-1}$ is the lower limit of
velocity dispersion set by the discrete nature of the SPH simulation,
while the upper limit of $800~{\rm km~s}^{-1}$ is determined by the
small scale turbulence in the debris. The velocity dispersion of
particles in the halo (about $1500$ particles in the 20k run) is
significantly higher and reaches $6000~{\rm km~s}^{-1}$. As the value
of the velocity dispersion increases the profile features get smoothed
out, until only a smooth, double-peaked profile is observed.

\item[{\it Effect of illuminating light curve. --}] We have computed 
model profiles for several different X-ray illumination light curves
keeping all the other parameters fixed. We used (a) the light curve
obtained from the accretion rate in the 5k simulation, (b) the light
curve from the accretion rate as predicted by theory, i.e.  $\propto
t^{-5/3}$ \citep{Rees, Phinney}, and (c) a light curve that is
constant in time (Figure~\ref{fig_lcurve}). We find that the line
profile shapes do not depend sensitively on the shape of the light
curve. This is a consequence of the centrally ''weighted'' emissivity
profile of the debris which causes the innermost emission region to be
the dominant contributor of the H$\alpha$ light. In the innermost
region of the debris the dynamic range in light-travel times is not
large; therefore the illumination of the innermost emitting region is
almost instantaneous. Over the very short light-crossing time of the
central emitting region, the gradient in the UV/X-ray light curve is
small and the illumination is nearly constant over this region. The
fast fluctuations in the illuminating light curve on the other hand
are smoothed out during reprocessing in the debris, and cannot be
identified in the H$\alpha$ light curve.

\end{description}

In summary, we find that profiles are not significantly influenced by
the shape of the illuminating light curve. The profile shapes are
affected, however, by the inner radius of the line-emitting region and
redistribution of the debris. The inner radius can change with the
advance or recession of the ionization front into the debris, which is
controlled primarily by the density of the debris. The most notable
effect of the inner radius is on the width of the profile. Since the
physical conditions can change very rapidly during flares, this
mechanism causes the line profiles to change on the light crossing
time scale (minutes to hours) and evolve from wide multi-peaked to
narrow and vice versa (the recombination time is negligible in
comparison to the light-crossing time of the debris, therefore
particles in the debris respond effectively instantaneously to changes
in the incident flux). The redistribution of the debris in phase
space, on the other hand takes more time: $\sim$24 hours for particles
in the innermost part of the emitting region. This redistribution of
the debris also may cause a transition from narrower to wider
profiles, however it takes place gradually, on the time scale of days.

\section{Discussion}

\subsection{Physical Conditions in the Debris and Radiative Processes\label{S_phys}}

We have calculated the physical conditions and radiative processes in
the debris using the photoionization code CLOUDY, version 94
\citep{Ferland}. It is necessary to know the physical conditions in
order to assess the validity of the assumptions made in our line
profile calculations.

Since the physical conditions in the tidal tail, disk and halo differ
noticeably, these three regions naturally emerge as separate
components of the tidal debris. The spheroidal halo is an oblate
structure of particles scattered out of the plane of the debris. The
disk is produced by the flow of particles from the tail which turn
around the black hole and form a higher density circular component
concentrated close to the plane of the debris. The disk shows a smooth
transition to the halo in terms of density and physical parameters
(see Table~\ref{T_debris}). The physical differences among the
components arise as a consequence of the number density, which is
several orders of magnitude higher in the tail. The particle density
of the halo is uncertain, due to the numerical scatter, and
consequently its luminosity contribution is also subject to
uncertainty. We nevertheless have calculated and presented physical
properties of the spheroidal component, as given by the SPH
simulation, and we discuss the implications of its presence for the
line profiles and the observability in the next section.

The temperature of the debris tail reaches $3\times10^{4}\,$K in the
hottest parts of the tail (i.e. the illuminated face of the tail) and
equals $5000\,$K on average in the partially ionized and neutral parts
of the tail. The temperature in the disk ranges from $8000\,$K in the
inner region to $1\times 10^{6}\,$K at the outer rim of the disk, with
the average $\sim 10^{5}\,$K. The mean temperature of the halo is
$1\times10^{4}\,$K and ranges from $3\times 10^{7}\,$K in the central
parts of halo, to $1\times 10^{4}\,$K in the outer halo. The
ionization parameter is calculated for all components as $U\equiv
\Phi_H/n_H\,c$, where $\Phi_H$ is the flux of ionizing
photons, $n_H$ is the total hydrogen density, and $c$ is the speed of
light. The ionization parameter in the tail ranges from 0.1 in the
parts of the tail closest to the source of ionization, to $10^{-5}$ on
the far side of the tail. The ionization parameter in the disk is
almost constant throughout the disk with a value of about 20.  The
ionization parameter in the halo is higher with an average value of
about 27. As a consequence, the disk and the halo are in a much higher
state of ionization relative to the tail. The disk shows a wide range
of hydrogen ionization fractions over its radius, with the strongest
ionization at the outer rim of the disk where density is lowest. In
the halo the fraction of neutral hydrogen atoms ranges from
$10^{-10}$ to $10^{-7}$, with an average value close to $10^{-7}$. In
contrast, the tail is mostly neutral, with only one hydrogen ion per
$10^{3}$ hydrogen nuclei. The properties of the tidal disruption
regions are summarized in Table~2.

The ionization state of the debris as well as its abundance and
density determine the radiative processes dominant in the debris. The
implication of the assumed solar metalicity is that emission lines
from metals play an important role in the cooling of the debris. We
focus our attention on processes relevant to the final H$\alpha$
luminosity. Two processes contribute to the H$\alpha$ luminosity of
the debris: recombination and collisional excitation. The dominant
recombination channel is the recombination of a photoelectron to the
$n=3$ level followed by a decay to the $n=2$ level via emission of an
H$\alpha$ photon. Emission by collisional excitation occurs when
hydrogen atoms in the $n=1$ and $n=2$ levels are promoted to the $n=3$
level by collisions with energetic photoelectrons and then de-excite
radiatively. The relative contribution of H$\alpha$ emission through
recombination relative to collisions is
\begin{equation}
\frac{L_{\rm H\alpha, rec}}{L_{\rm H\alpha, coll}} =
\frac{\alpha_{rec}}{\alpha_{coll}}\frac{ n_{HII}}{n_{HI}}\frac{V_{rec}}{V_{coll}}
\label{eq_ratio}
\end{equation}
where $\alpha_{rec}$ and $\alpha_{coll}$ are the respective
coefficients for recombination and collision processes which lead to
emission of an H$\alpha$ photon, $n_{HII}$ and $n_{HI}$ are the number
density of hydrogen ions and atoms, and $V_{rec}$ and $V_{coll}$ are
the parts of the debris volume which gives rise to emission through
recombination and collisional de-excitation, respectively.

The recombination coefficient for an electron to recombine from the
continuum to the n=3 level can be writen as $\alpha_{C} =
\sum_{n=3}^{\infty} \alpha_{n}$ for a given temperature. The adopted
recombination coefficients for the tail and halo are
$\alpha_{rec}^{tail} = 3.4 \times 10^{-13}~{\rm cm^{3}~s^{-1}}$,
$\alpha_{rec}^{disk} \simeq 1.8 \times 10^{-14}~{\rm cm^{3}~s^{-1}}$, and
$\alpha_{rec}^{halo} = 1.8 \times 10^{-13}~{\rm cm^{3}~s^{-1}}$
\citep{Osterbrock}. Similarly, it is possible to estimate the rate of 
collisions which will promote an electron to n=3 state. The high
optical depths in the Lyman series cause a significant electron
population in the $n=2$ level. The photoelectron energy distribution
for low-energy levels of hydrogen differs from the Boltzmann
distribution and the number density is not sufficiently high for
collisions to lead to thermal equilibrium \citep[Eq. 10.57
of][]{Krolik}. In particular, thermalized energy levels will be n$>$1
for the tail, n$>$2 for the halo, and n$>$3 for the disk, and the
relative populations of these levels will be in agreement with their
statistical weights. It is possible then to estimate the collisional
excitation coefficients for transitions from the n=1 and n=2 levels to
the thermalized n=3 level
\citep[Eq. 3.21 of][]{Osterbrock}. This approximately yields
collisional excitation rates for the tail, disk and halo of
$\alpha_{coll}^{tail} = 1.2 \times 10^{-11}~{\rm cm^{3}~s^{-1}}$, 
$\alpha_{coll}^{disk} = 1.1 \times 10^{-9}~{\rm cm^{3}~s^{-1}}$, and
$\alpha_{coll}^{halo} = 9.0 \times 10^{-11}~{\rm cm^{3}~s^{-1}}$.

With the values of the rate coefficients, the ionization state of the
tail, disk and halo ($n_{HII}/n_{H}\approx 10^{-3}$ in the tail and
$10^{7}$ in the halo and disk on average), and taking into account the
ratio of emitting volumes for two processes, it is possible to
calculate the ratio of respective relative contributions from
recombination and collisions from equation~(\ref{eq_ratio}) to be
about $1\times10^{-6}$, $6\times10^{3}$, and $2\times10^{4}$. These
values are not unexpected: collisional excitation is the dominant
mechanism for production of H$\alpha$ photons in the tail, where the
density is highest, while recombination is dominant in the disk and
halo.

Some of the H$\alpha$ photons created by the two processes are
destroyed on the way out of the debris. The mechanisms that impact the
total output in H$\alpha$ luminosity are the absorption in the
H$\alpha$ transition, and electron scattering, with respective optical
depths $\tau_{H\alpha}$ and $\tau_e$. The optical depth
$\tau_{H\alpha}^{tail}$ is very high for the H$\alpha$ photons
traveling across the tail, along the lines of incidence of
illuminating photons and typically has a value of order
$10^{7}$. Electron scattering in the tail has an average optical depth
of $\tau_{e}^{tail}\approx 10$. Consequently, the majority of
H$\alpha$ photons are created in the thin ionized and partially
ionized layers of the tail (we refer to them as the tail
``skin''). From here a certain fraction of H$\alpha$ photons escapes
the debris and reaches the observer. In the disk
$\tau_{H\alpha}^{disk}\approx 4\times10^{6}$ and
$\tau_{e}^{disk}\approx 8$. In the early stages of accretion, while
the luminosity is still super-Eddington, the halo is optically thin to
H$\alpha$ photons ($\tau_{H\alpha}^{halo}\approx 2\times10^{-2}$) and
the electron scattering optical depth is about $\tau_{e}^{halo}\approx
5$. In the late stages of the debris evolution
$\tau_{H\alpha}^{halo}\approx 1\times10^{4}$ and $\tau_{e}$ stays
approximately the same in the halo. The high values of $\tau$ of the
halo material destroy the fraction of the H$\alpha$ photons emitted by
the part of the debris embedded in the halo. The evolution of the
spherical halo would further help the process of fading of H$\alpha$
emission with time along with the decreasing accretion luminosity. For
the purposes of radiative transfer calculations we have modeled the
tail, the disk and the halo as three {\it separate} components and
therefore we do not account for the secondary absorption and electron
scattering by the halo, of H$\alpha$ photons created in the tail. Both
opacity mechanisms cause the destruction of H$\alpha$ photons, if the
spherical halo is fully evolved, and possibly wipe out the H$\alpha$
emission line. If the same process operates in the centers of galaxies
which are tidal disruption candidates, it should be possible to
observe the disappearance of the broad H$\alpha$ emission line on the
scale of months.

\subsection{Spheroidal Halo: Implications for Line Profiles and
Luminosity\label{S_halo}}

In our predictions of the observational signature of the
post-disruption debris we have not included any contributions from the
spheroidal halo because we doubt that this structure, as predicted by
our SPH simulation is real. In this section we discuss this issue
further and justify our approach. We also describe qualitatively the effect 
of such a halo on the observational appearance of the debris.

The halo is very likely produced artificially because a number of
particles is scattered from the disk every time the flow of particles
turns around the black hole and intersects with the inflowing stream
of particles.  The number of particles contributed to the halo by
numerical scatter is proportional to $\sqrt{N}$, where N the total
number of particles in the run, and multiplied by a factor
$\tau_{run}/\tau_{dyn}$, as described in \S2. The dynamical
time scale of the disk is about 7 days for the outermost particles,
which implies that the number of scattered particles is of the order
of 1000. This is an upper limit of the uncertainty, where total number
of particles in the halo, in the 20k run, reaches number of 1500
particles in the later stages of the debris evolution.

It has been pointed out by several authors that a spheroidal halo
\citep{UPG,LU,CLG,AL} with a central radiatively supported torus
\citep{Rees,LU,CLG,EK} may form around the black hole as a consequence
of a self-compression of the flow, which leads to the ejection of the
debris with high velocities, perpendicular to the plane. However, none
of the above mentioned structures were resolved beyond doubt in the
simulations. A halo-like structure is expected to be in hydrostatic
equilibrium, and its size should be determined by the radiation
pressure from the central source. \citet{UPG} and \citet{LU} argue
that the luminosity incident upon the halo can rarely be exactly tuned
for the structure to rest in equilibrium. They further calculate that
for a super-Eddington luminosity the halo will expand and cool down on
the time scale of months to years, until it becomes gravitationally
unbound and is blown away. For a sub-Eddington luminosity, on the
other hand, the halo is expected to collapse due to insufficient
support from radiation pressure.

Based on its physical conditions it is obvious that, if real, the halo
observed in our simulations would make the dominant contribution to
the line emission from the debris. In the late stages of the debris
evolution, our simulations show the halo becoming optically thick to
H$\alpha$ photons.  Moreover, the long diffusion time of photons in
the halo ($t_{diff}\sim\tau_{e} \; r/c\approx3.5$ hours, where $r$ the
is inner radius of the halo) may smear out the time variability of the
emission line profiles, which is one of the main signatures of the
disruption event.  The high value of the velocity dispersion of the
particles in the halo would smear out the line profiles and very
likely make them unobservable. It is therefore necessary to address in
future studies what fraction of the halo (if any) is really present
and what fraction is contributed by the numerical scatter.

If a halo manages to form, survives the super-Eddington phase, and
achieves hydrostatic equilibrium it will be transparent to H$\alpha$
photons only during the super-Eddington phase. At later times, we
expect that it will become optically thick and will be the primary
source of illumination of the outer debris.  Moreover, it will fade on
a time scale of months to years. In this picture, which is consistent
with the predictions of \citet{LU}, the shape of the observed line
profiles should be similar to the profiles computed here for large
values of the inner radius (see Figure~\ref{fig_rad}). Whether the
profile (noticeable here for $\xi=10,000$ at the wavelength
$\lambda=6500\AA$) will appear as blueshifted or redshifted would
depend on the orientation of the debris tail with respect to the
observer.

\subsection{Observability of Emission Lines and Their Uniqueness as a
Tidal Disruption Signature\label{S_obs}}

The CLOUDY calculation predicts a time-average H$\alpha$ luminosity
from the tidal debris of about $10^{36}$, $10^{37}$, and $6\times
10^{38}~{\rm erg~s^{-1}}$ for the tail, disk and halo,
respectively. In the earlier stages of the disruption event when the
UV/X-ray luminosity is super-Eddington, the H$\alpha$ luminosity is
expected to be up to 80 times higher than its average value and
comparable to that of tidal disruption candidates observed in the
local universe (see Figure~\ref{fig_lcurve}). The examples are
NGC~4450 (at 16.8~Mpc) with an H$\alpha$ luminosity of
$L_{H\alpha}$=1.8$\times 10^{39}~{\rm erg~s^{-1}}$ \citep{Ho} and
NGC~1097 (at 22~Mpc) with $L_{H\alpha}$=7.7$\times 10^{39}~{\rm
erg~s^{-1}}$ \citep{SB95}. Thus the emission-line signature of a tidal
disruption event should be detectable at least out to the distance of
the Virgo cluster. In practice, however, the detection of such
emission lines from low luminosity sources may be complicated by the
weak contrast relative to the underlying stellar continuum.

Based on observational constraints from known tidal disruption
candidates, it should be possible to detect some variable properties
of the line profiles and light curves predicted here. One of the first
observable effects of tidal disruption should be UV/X-ray flash
accompanied by a decaying light curve, mirrored in the delayed
response of the H$\alpha$ light curve of the debris, with some
scatter. The line profile intensities are expected to decay
accordingly in time. Other effects to look for are the change in the
number of peaks in the line profile, relative fluctuations
in intensity of the peaks as well as their shift in wavelength.

The temporal variability of the H$\alpha$ emission line profiles from
the post-disruption debris is one of the important indicators of a
tidal disruption event. In order to capture the rapid profile
variability, due to the variable illumination, the exposure time
should be comparable to the light crossing time of the innermost
regions of the line-emitting debris, which has the fastest and
strongest response to the ionizing radiation. Longer exposures are
expected to capture the average shape of the rapidly varying line
profiles. The light-crossing time of the innermost regions of the
debris is about $8\;\xi_{2}\;M_{6}$~minutes (where $\xi_{2}=\xi/100$),
while the exposure times are typically about 30-60 minutes (for
galaxies at the distance of the Virgo cluster, for example). Thus, if
an event is caught early in its evolution and the light-crossing time
is relatively long (i.e., $M \gtrsim 10^{6}~\Msun$), there is a chance
of detecting variability caused by changing illumination over the
course of one to a few nights. On longer time scales, variability is
caused by changes in the structure of the debris. In the presence of
the spheroidal halo, the variability of the lines may be modified by
the long diffusion time scale of photons through the halo. The
component of the tidal tail outside the halo will then still respond
to the variability but on the time scale set by the light reprocessed
by the halo.

In view of the predictions from our profile calculation, an important
question is whether variable multi-peaked line profiles can originate
in some other physical scenario or can be regarded as the unique
signature of tidal disruption. Multi-peaked emission lines are likely
to be the signature of an inhomogeneity in the phase-space
distribution of the emitting material. Due to the asymmetry of the
emitting region, the direction of the observer has a large influence
on the observed line profile, making it difficult, if not impossible,
to infer the exact emission geometry from a particular multi-peaked
profile. Nevertheless the variability pattern of the line profiles can
serve as a general indicator of a tidal disruption event.  Additional
observational indicators can be used in conjunction to diagnose a
tidal disruption event, for example, a sharp X-ray/UV flash preceding
the appearance of the emission lines, an emission line spectrum
indicative of a hard ionizing continuum (i.e., the presence of ionic
species with high ionization potentials), and the characteristic decay
of the emission-line flux on time scales of weeks to months after the
event \citep*[the flux of different emission lines is expected to
decay at different rates; see, for example,][]{ELB95}.

\subsection{Approximations in the Calculation \label{S_approx}}

The finite resolution of the simulation could introduce uncertainties
in the accretion light curve caused by the discretized accretion of
the stellar material onto the black hole. Another possible effect is
that the finite number of particles does not completely reveal the
morphology of the debris and some of its components may stay
hidden. For example, when the majority of the particles are confined
to the tail it is hard to say if there is an accretion disk forming
around the black hole out of a small number of
particles. Consequentially, there is a concern that profiles
calculated for low values of the inner radius, i.e. $\xi_{in}$=200,
may represent the contribution from a small number of particles in the
central region and therefore dominated by small-number noise.

A possible source of error is the assumption of a {\it thin, flat}
structure (i.e., confined to a plane). In the early stages of the
evolution of the debris, the majority of the particles are still in
the tidal tail, located in a plane, which makes the assumption
valid. During the evolution of the debris the number of particles that
orbit around the black hole in an almost spherical distribution
increases \citep{CLG,LU,UPG,Ulmer,MQ}. At that point it is possible to
distinguish three structural components of the debris: the relatively
planar tidal tail and the disk and the spheroidal halo. The halo is
made up of particles scattered from the tail by shocks during the
pericentric approach of the debris or during the intersection of the
tail with itself \citep{Kochanek,LKR,KPL,ALP} and some fraction of
particles contributed by the numerical noise. Since the tidal tail
includes the majority of the particles, and most of the halo mass is
concentrated close to the equatorial plane, even in our last frame,
the assumption of a thin disk is still reasonable. If, however, the
mass in the spherical halo increases at very late times, the
assumption of a planar geometry needs to be reconsidered.

We also adopt the weak field approximation in our calculations, which
is a fairly small source of error (of order 1\% and less). This is
valid since we adopt $\xi_{in}=100$ as the innermost radius of the
line-emitting debris. The few particles interior to this radius would
not contribute to the H$\alpha$ emission because their close proximity
to the source of ionizing radiation would make them fully ionized. In
the case of a physical scenario where emission of H$\alpha$ is not
possible because of a highly ionized debris, the same profile
formalism can be used to calculate the emission from other lines
emitted under these conditions.

\section{Conclusions}

We modeled the emission-line luminosity and profile from the debris
released by the tidal disruption of a star by a black hole in the
early phase of evolution. Our model predicts prompt optical evolution
of post-disruption debris and profile shapes different from circular
and elliptical disk model profiles. Since line profiles observed so
far in LINERS look more disk-like and evolve slowly, the observations
are likely to have caught the event at late times ($\ge$ 6 months
after the initial disruption), after the debris has settled into a
quasi-stable configuration.

The line profiles can take a variety of shapes for different
orientations of the debris tail relative to the observer. Due to the
very diverse morphology of the debris, it is almost impossible to
uniquely match the multi-peaked profile with the exact emission
geometry. Nevertheless, the profile widths and shifts are strongly
indicative of the velocity distribution and the location of matter
emitting the bulk of the H$\alpha$ light. Profile shapes do not depend
sensitively on the shape of the light curve of the X-rays illuminating
the debris. They strongly depend on the distance of the emitting
material from the central ionizing source, which is a consequence of
the finite propagation time of the ionization front and the
redistribution of the debris in phase space. It may be possible to
distinguish between the two effects observationally, based on their
different characteristic time scales. The onset of the optically thick
spheroidal halo should cause the disappearance of the broad H$\alpha$
emission line on the time scale of months, and give rise to the
emission of narrower, strong, blueshifted or redshifted emission line,
arising from the portion of the tidal tail unobscured by the halo.

If X-ray flares and the predicted variable profiles could be observed
from the same object they could be used to identify the tidal
disruption event in its early phase. The X-ray flares can be promptly
detected by all-sky synoptic X-ray surveys and high energy burst alert
missions such as {\it Swift}. The evolution of the tidal event may
then be followed with optical telescopes from the ground on longer
time scales and give an insight in the next stage of development of
the debris. Thus, simulations of the tidal disruption process on
longer time scales (of order several months to a few years) are sorely
needed. Calculations of the long-term evolution of a tidal disruption
event can predict the type of structure that the debris finally
settles into and whether its emission-line signature resembles the
transient double-peaked lines observed in LINERs. This study would
provide an important insight into the evolution of LINERs.

Finally, the observed rate of tidally disrupted solar type stars can
constrain the rate of captured compact objects (which are important
gravitational wave sources), and the capture rate of main sequence
stars in our Galaxy, which are expected to emit the peak of the
gravitational radiation in the LISA frequency band and can be detected
in the local universe.

\acknowledgements

We are indebted to J. Charlton for her help with CLOUDY. T.B. also
thanks M. Falanga for his valuable comments. We are also grateful to
the anonymous referee for very insightful and helpful comments and
suggestions. We acknowledge the support of the Center for
Gravitational Wave Physics funded by the NSF under cooperative
agreement PHY-0114375, NSF grants PHY-9800973 and PHY-0244788, the
Zaccheus Daniel Fellowship, and the Eberly College of Science.

%%%%%%%%%%%%%%%%%%%%%%%%%%%%%%%%%%%%%%%%%%%%%%%%%%%%%%%%%%%%%%%%%%%%%
%%%%%%% R E F E R E N C E S
%%%%%%%%%%%%%%%%%%%%%%%%%%%%%%%%%%%%%%%%%%%%%%%%%%%%%%%%%%%%%%%%%%%%%

%%%%%%%%%%%%%%%%%%%%%%%%%%%%%%%%%%%%%%%%%%%%%%%%%%%%%%%%%%%%%%%%%%%%%
%%%%%%% T A B L E S
%%%%%%%%%%%%%%%%%%%%%%%%%%%%%%%%%%%%%%%%%%%%%%%%%%%%%%%%%%%%%%%%%%%%%
\clearpage

\begin{deluxetable}{ccccccc} 
\tablecaption{Parameters of SPH Runs\label{T_sph}}
\tablewidth{0pt}
\tablecolumns{7}
\tablehead{
\colhead{} & 
\colhead{} & 
\colhead{Total} & 
\colhead{Start of} & 
\colhead{Emission-Line} & 
\colhead{Time} & 
\colhead{Particle} \\
\colhead{SPH} & 
\colhead{Total} & 
\colhead{Duration} & 
\colhead{Accretion} & 
\colhead{Evolution} & 
\colhead{Step} & 
\colhead{Mass} \\
\colhead{Run} & 
\colhead{Particles} & 
\colhead{(days)} & 
\colhead{(days)} & 
\colhead{(days)} & 
\colhead{(hours)} & 
\colhead{(g)}
}
\startdata
 5k & 5000 & 94 & 34 & 60 & 4.11 & 3.978$\times 10^{29}$\\ 
 20k & 20000 & 53 & 47 & 6 & 0.746 & 9.945$\times 10^{28}$\\ 
\enddata
\end{deluxetable}

\begin{deluxetable}{lccccccc} 
\tablecaption{Physical Properties of the Debris\label{T_debris}}
\tablewidth{0pt}
\tablecolumns{8}
\tablehead{
\colhead{} &
\colhead{$N_{H}$ \tablenotemark{\rm a}} & 
\colhead{$n_{H}$} & 
\colhead{$n_{HI}/n_{Htot}$} & 
\colhead{$n_{HII}/n_{Htot}$} & 
\colhead{$T$ \tablenotemark{\rm b}} &
\colhead{U \tablenotemark{\rm c}} & 
\colhead{$L_{\rm H\alpha}$ \tablenotemark{\rm d}} \\
\colhead{Debris Region} & 
\colhead{(cm$^{-2}$)} & 
\colhead{(cm$^{-3}$)} & 
\colhead{} & 
\colhead{} & 
\colhead{(K)} & 
\colhead{} &
\colhead{(erg s$^{-1}$)}
}
\startdata
 Tail            & $10^{27}$--$10^{31}$ & $10^{14}$--$10^{15}$ & $\sim1$        &  
$\sim 10^{-3}$ & $5\times 10^{3}$ & $10^{-5}$--0.1 &  $1\times 10^{36}$ \\ 

 Disk		 & $10^{21}$--$10^{25}$ & $10^{11}$--$10^{12}$ & $10^{-8}$--$0.3$ &
$0.7$--$1$     & $\sim1\times 10^{5}$ & 20 & $1\times 10^{37}$ \\ 

 Halo            & $10^{25}$            & $10^{9}$--$10^{12}$  & $\sim 10^{-7}$ & 
$\sim1$       & $1\times 10^{4}$ & 27             &  $6\times 10^{38}$ \\ 

\enddata
\tablenotetext{a}{The column density: low
  and high values correspond to directions orthogonally and radially
  through the debris component.}
\tablenotetext{b}{The average value of temperature over radius.}
\tablenotetext{c}{Ionization parameter: range of values in the tail and average in the disk and halo.}
\tablenotetext{d}{The H$\alpha$ luminosities from the debris components as calculated for the time-average value of incident luminosity $L_{acc}=1.5\times 10^{43}$erg s$^{-1}$.}
\end{deluxetable}

%%%%%%%%%%%%%%%%%%%%%%%%%%%%%%%%%%%%%%%%%%%%%%%%%%%%%%%%%%%%%%%%%%%%%
%%%%%%% F I G U R E S
%%%%%%%%%%%%%%%%%%%%%%%%%%%%%%%%%%%%%%%%%%%%%%%%%%%%%%%%%%%%%%%%%%%%%

\clearpage

\clearpage

\end{document}